\documentclass{article}

\usepackage[preprint]{neurips_2025}

\usepackage[utf8]{inputenc} % allow utf-8 input
\usepackage[T1]{fontenc}    % use 8-bit T1 fonts
\usepackage{hyperref}       % hyperlinks
\usepackage{url}            % simple URL typesetting
\usepackage{booktabs}       % professional-quality tables
\usepackage{amsfonts}       % blackboard math symbols
\usepackage{nicefrac}       % compact symbols for 1/2, etc.
\usepackage{microtype}      % microtypography
\usepackage{xcolor}         % colors

\usepackage{fancyhdr}       % header
\usepackage{graphicx}       % graphics
\usepackage{tikz}
\usepackage{pgfplots}
\usepgfplotslibrary{polar}
\pgfplotsset{compat=1.18}
\usepackage{pgf-pie}  % For pie charts
\usepackage{subcaption}
\usepackage{amssymb} % Added for additional math symbols
\usepackage{pgfplotstable} % For reading CSV files
\usepackage{multirow} % For multirow tables
\usepackage{listings} % For code listings

\definecolor{python_blue}{RGB}{0, 114, 178}
\definecolor{cpp_orange}{RGB}{230, 159, 0}

\pgfplotstableread[col sep=comma]{python_gemini-2.5-pro-preview-03-25_violin.csv}\pythongeminitwodata
\pgfplotstableread[col sep=comma]{cpp_gemini-2.5-pro-preview-03-25_violin.csv}\cppgeminitwodata
\pgfplotstableread[col sep=comma]{python_deepseek-ai_DeepSeek-V3_violin.csv}\pythondeepseekdata
\pgfplotstableread[col sep=comma]{cpp_deepseek-ai_DeepSeek-V3_violin.csv}\cppdeepseekdata
\pgfplotstableread[col sep=comma]{python_deepseek-reasoner_violin.csv}\pythondeepseekreasonordata
\pgfplotstableread[col sep=comma]{cpp_deepseek-reasoner_violin.csv}\cppdeepseekreasonordata
\pgfplotstableread[col sep=comma]{python_gemini-2.5-flash-preview-04-17_violin.csv}\pythongeminiflashdata
\pgfplotstableread[col sep=comma]{cpp_gemini-2.5-flash-preview-04-17_violin.csv}\cppgeminiflashdata
\pgfplotstableread[col sep=comma]{python_openrouter_google_gemma-3-27b-it_violin.csv}\pythongemmarouter
\pgfplotstableread[col sep=comma]{cpp_openrouter_google_gemma-3-27b-it_violin.csv}\cppgemmarouter
\pgfplotstableread[col sep=comma]{python_meta-llama_Llama-4-Maverick-17B-128E-Instruct-FP8_violin.csv}\pythonllamamaverick
\pgfplotstableread[col sep=comma]{cpp_meta-llama_Llama-4-Maverick-17B-128E-Instruct-FP8_violin.csv}\cppllamamaverick
\pgfplotstableread[col sep=comma]{python_meta-llama_Llama-4-Scout-17B-16E-Instruct_violin.csv}\pythonllamascout
\pgfplotstableread[col sep=comma]{cpp_meta-llama_Llama-4-Scout-17B-16E-Instruct_violin.csv}\cppllamascout
\pgfplotstableread[col sep=comma]{python_Qwen_Qwen3-235B-A22B-fp8-tput_violin.csv}\pythonqwendata
\pgfplotstableread[col sep=comma]{cpp_Qwen_Qwen3-235B-A22B-fp8-tput_violin.csv}\cppqwendata
\pgfplotstableread[col sep=comma]{python_deepseek-ai_DeepSeek-R1-Distill-Llama-70B_violin.csv}\pythondeepseekdistilldata
\pgfplotstableread[col sep=comma]{cpp_deepseek-ai_DeepSeek-R1-Distill-Llama-70B_violin.csv}\cppdeepseekdistilldata

\title{CodeGolf Bench: A Multi-Language Benchmark for Evaluating Concise Code Generation Capabilities of Large Language Models}

\author{%
  Vedant Padwal \\
  Independent \\
  \texttt{vedantpadwalinfi@gmail.com} \\
}

\date{May 2025}

\begin{document}

\maketitle
\vspace{-0.3in}
\begin{center}
\textit{16 May 2025}
\end{center}
\vspace{0.1in}

\begin{abstract}
This paper introduces Code Bench, a benchmark capable of evaluating Large Language Models (LLMs) concise code generation abilities in 60 programming languages. Based on code golf, a recreational programming competition focused on minimal character or byte solutions, the benchmark provides a distinctive measure of LLMs ability to produce efficient, concise code. Unlike existing benchmarks limited by fixed problem sets and language coverage, CodeGolf Bench leverages the code.golf platform to provide new problems and live human performance baselines. Evaluation of nine LLMs on Python and C++ tasks demonstrates that reasoning models significantly outperform non-reasoning models, achieving best average percentile of 70.97\%. This performance gap is particularly pronounced in C++, highlighting reasoning's importance for languages with strict syntax requirements. Non-reasoning models struggle more with efficiency optimization across both languages, with best percentiles significantly lower than reasoning counterparts. CodeGolf Bench offers a dynamic framework for evaluating LLM code generation capabilities against evolving human performance on code golf.\end{abstract}
\section{Introduction}
Code golf is a recreational programming competition where participants solve problems using the fewest characters possible \cite{wiki:codegolf}, originating in the 1990s Perl community and expanding to various languages, including specialized golfing dialects like GolfScript. It serves as a unique metric for evaluating language expressiveness and programmer ingenuity, often producing creative yet compact solutions. The focus on minimizing code size is particularly relevant for Internet of Things (IoT) development, where devices face memory, processing, and power constraints~\cite{rozlomii2024data, beltran2024review}. Techniques honed in code golf, such as writing concise and efficient code, can enhance skills applicable to optimizing resource-constrained systems like IoT devices~\cite{bungo2008fundamental}. Though primarily recreational, code golf’s emphasis on efficiency informs practical applications in embedded systems programming, where resource limitations are a critical concern~\cite{10.5555/517270}.

Code generation using Large Language Models (LLMs) has witnessed remarkable progress in recent years, particularly accelerated by the emergence of reasoning LLMs \cite{deepseekai2025deepseekr1incentivizingreasoningcapability}, which have rapidly established themselves as the state of the art in coding benchmarks \cite{chen2021evaluating, hendrycks2021measuring, austin2021program}. While traditional non-reasoning models primarily generate code based on patterns observed in their training data often struggling with novel problems introduced after their training cutoff date reasoning models demonstrate superior ability to understand problem statements and synthesize optimal solutions through step-by-step logical deduction \cite{white2025livebenchchallengingcontaminationlimitedllm, chambon2025bigobenchllmsgenerate}.

Despite these advancements, current code generation benchmarks face several critical limitations. Most of them are manually curated with fixed set of problems that quickly become saturated as model performance improves \cite{shi2024languagemodelssolveolympiad}. Additionally, these benchmarks predominantly focus on widely used languages such as Python, C++, and Java, leaving a significant gap in the evaluation of LLM performance across the diverse programming language ecosystem \cite{huang2025effibenchbenchmarkingefficiencyautomatically, yu2024humanevalprombpppro}.

Furthermore, existing benchmarks often lack robust human baselines, instead evaluating models only in relation to other LLMs. While recent work like CodeElo \cite{quan2025codeelobenchmarkingcompetitionlevelcode} has attempted to address this by incorporating human performance metrics, these measurements represent static snapshots that fail to capture the evolving capabilities of the human programmers.

To address these limitations, this paper presents CodeGolf Bench, a novel benchmark that has the ability to evaluate LLM code generation capabilities across 60 different programming languages substantially broader than any existing benchmark. The approach leverages the \url{code.golf} platform to provide continuously updated problems and, crucially, live human baselines that reflect real-time human performance. This dynamic comparison allows for more meaningful assessment of LLM coding abilities in the context of human expertise, particularly within the challenging constraint optimization paradigm that code golf represents.

The CodeGolf Bench includes diverse problem categories ranging from visual art and gaming to mathematical sequences and text transformation tasks. Through comprehensive evaluation across these domains, this research demonstrates that reasoning capabilities significantly enhance model performance in coding tasks, particularly for languages with strict syntax requirements like C++. The findings highlight the importance of reasoning abilities in producing not only correct but also efficient solutions across programming paradigms.
\section{Related work}
\label{gen_inst}

\subsection{Code-generation benchmarks}
\label{code_gen_benchmarks}

When benchmarks like HumanEval~\cite{chen2021evaluating}, APPS~\cite{hendrycks2021measuring}, and MBPP~\cite{austin2021program} were released, the coding capabilities of LLMs were limited to simple tasks. As 
their capabilities improved and these benchmarks got saturated, researchers designed harder benchmarks like xCodeEval~\cite{khan-etal-2024-xcodeeval} is based on questions from competitive programming websites.
To address the issue of contamination, LiveBench~\cite{white2025livebenchchallengingcontaminationlimitedllm} is designed to collect new problems from LeetCode, AtCoder, and Codeforces.

BigO(Bench)~\cite{chambon2025bigobenchllmsgenerate}, EffiBench~\cite{huang2025effibenchbenchmarkingefficiencyautomatically} and COFFE~\cite{peng2025coffe} benchmarks are designed to evaluate the capabilities of
generative language models in understanding and generating code with specified time and space complexities. Benchmarks like SWE-Bench~\cite{jimenez2023swebench} and Aider Polyglot~\cite{aiderpolyglot2025} are designed to test agentic coding abilities of LLMs on real-world software engineering tasks.

While these benchmarks are useful for evaluating different capabilities of LLMs, they typically use the pass@k metric and scores are measured relative to other LLMs. Most benchmarks focus on a limited set of programming languages, with Python being the most common and do not provide comprehensive evaluation across diverse programming languages. As demonstrated in Table~\ref{tab:benchmark}, even the most language-diverse benchmark before CodeGolf Bench (xCodeEval~\cite{khan-etal-2024-xcodeeval}) only supports 11 languages, while CodeGolf Bench substantially expands this coverage to 60 languages.

\subsubsection{Competitive programming benchmarks}

Table~\ref{tab:benchmark} provides a comprehensive comparison of existing code generation benchmarks. CodeElo~\cite{quan2025codeelobenchmarkingcompetitionlevelcode} specifically targets competitive programming with Codeforces contests. The metrics used in CodeElo~\cite{quan2025codeelobenchmarkingcompetitionlevelcode} and EffiBench~\cite{huang2025effibenchbenchmarkingefficiencyautomatically} are derived from the human performance at a specific point in time. This distribution naturally changes as individual human participants improve their skills and competition difficulty evolves over time.

\clearpage
\begin{table}
    \caption{Benchmark comparison}
    \label{tab:benchmark}
    \centering
    \small
    \begin{tabular}{p{2cm}p{2cm}p{4.75cm}p{1.25cm}p{2cm}}
    \toprule
    \textbf{Benchmark} & \textbf{Languages} & \textbf{Sources and Problem Count} & \textbf{Human-Baseline} & \textbf{Updates} \\ 
    \midrule
    APPS~\cite{hendrycks2021measuring}  
      & Python  
      & 10000 problems (5000 train + 5000 test); 131836 unit tests; 232444 human solutions; scraped from Codewars, AtCoder, Kattis, and Codeforces  
      & No  
      & Static (since May 2021)  
    \\ \hline

    HumanEval~\cite{chen2021evaluating}  
      & Python  
      & 164 hand-crafted programming tasks with docstrings and unit tests  
      & No  
      & Static (since June 2021)  
    \\ \hline

    MBPP~\cite{austin2021program}  
      & Python  
      & MBPP: 974 crowdsourced Python tasks; MathQA-Python: 23914 converted math problems  
      & No  
      & Static (since May 2021)  
    \\ \hline

    xCodeEval~\cite{khan-etal-2024-xcodeeval}  
      & 11 languages
      & $\sim$7.5K unique problems; 25M document-level examples (16.5B tokens); ExecEval execution-based evaluation  
      & No  
      & Static (since Jan 2024)  
    \\ \hline

    USACO~\cite{shi2024languagemodelssolveolympiad}  
      & C, C++, Java, Pascal, Python  
      & 307 past USACO contest problems with hidden test suites, reference solutions, and official analyses  
      & No  
      & Static (since Apr 2024)  
    \\ \hline

    LiveBench~\cite{white2025livebenchchallengingcontaminationlimitedllm}  
      & Python  
      & Continuously collects new problems from LeetCode, AtCoder, and Codeforces; contamination‑free splits  
      & No  
      & Active (since Jun 2024)  
    \\ \hline

    EffiBench~\cite{huang2025effibenchbenchmarkingefficiencyautomatically}  
      & Python  
      & 1000 efficiency‑critical LeetCode problems; canonical human solutions; stress‑test inputs  
      & Yes  
      & Static (since Oct 2024)  
    \\ \hline

    HumanEval Pro \& MBPP Pro~\cite{yu2024humanevalprombpppro}  
      & Python  
      & Expanded "Pro" extensions: 164 HumanEval Pro + 378 MBPP Pro tasks with canonical solutions  
      & No  
      & Static (since Dec 2024)  
    \\ \hline

    CodeElo~\cite{quan2025codeelobenchmarkingcompetitionlevelcode}  
      & C++, Python  
      & Codeforces contests scraped over last 6 months; pipeline submits to CF judge and computes human-aligned Elo ratings  
      & Yes  
      & Static (since Jan 2025)  
    \\ \hline

    COFFE~\cite{peng2025coffe}  
      & Python  
      & 398 function-level + 358 file-level tasks with contract-based "stressful" tests; \texttt{efficient@k} metric  
      & No  
      & Static (since Feb 2025)  
    \\ \hline

    BigO(Bench) \\
    ~\cite{chambon2025bigobenchllmsgenerate}  
      & Python  
      & 3105 contest problems + $\sim$1.19M solutions profiled for Big‑O labels; profiling/regression toolkit  
      & No  
      & Static (since Mar 2025)  
    \\ \hline

    CodeGolfBench
  & 60 Languages  
  & 115 Problems per language
  & Yes  
  & Live
  \\ \bottomrule
    \end{tabular}
\end{table}

As shown in Table~\ref{tab:benchmark}, CodeGolf Bench addresses limitations of existing benchmarks by supporting 60 different programming languages, significantly broader than any existing benchmark. Additionally, CodeGolf Bench uniquely provides live human baselines that continuously update to reflect the evolving capabilities of human programmers, allowing for more accurate and up-to-date comparisons between LLMs and human performance. Furthermore, unlike most static benchmarks that remain unchanged after publication, CodeGolf Bench continuously updates with new problems and scores, ensuring its relevance over time.

Table \ref{tab:benchmark_comparison} illustrates a key advantage of CodeGolf Bench: while most existing benchmarks have publicly available solutions, creating potential contamination risks, all submissions to \url{code.golf} remain private unless the submitter explicitly makes them public, significantly reducing contamination risk.

\clearpage
\section{CodeGolf Bench}
\label{headings}

CodeGolf Bench is based around the Code Golf platform. This section will cover details regarding the benchmark construction, Problem Categories, Solution Generation, Evaluation and Metrics Calculation.
In CodeGolf problems are referred to as holes, but they will be referred to as problems in this paper.

\subsection{Benchmark construction}

CodeGolf Bench leverages the established \url{code.golf} platform's extensive repository of programming challenges and validation infrastructure. The benchmark consists of three primary components: (1) a diverse set of programming challenges, (2) support for 60 programming languages, and (3) an evaluation framework that measures both correctness and conciseness.

To construct CodeGolf Bench, all available problems and supported languages from the \url{code.golf} platform are retrieved using its public API endpoints (\texttt{/holes} for problems and \texttt{/languages} for languages). Each problem includes a description, sample inputs/outputs, and a set of hidden test cases for validation. For each language, the \url{code.golf} platform supports automated execution and validation, resulting in a final set of 115 problems and 60 languages. All problem statements and metadata were programmatically retrieved and stored in a structured format to facilitate large-scale, automated benchmarking. This preprocessing step ensures that the benchmark remains up-to-date with the evolving \url{code.golf} ecosystem and can be easily extended as new problems or languages are added to the platform.

CodeGolf Bench also calculates comparative metrics against the distribution of human-written solutions, providing percentile rankings that contextualize model performance within the broader code golf community. Users can use the API of \url{code.golf} or setup their own server to host the platform.

\subsubsection{Problem Categories}

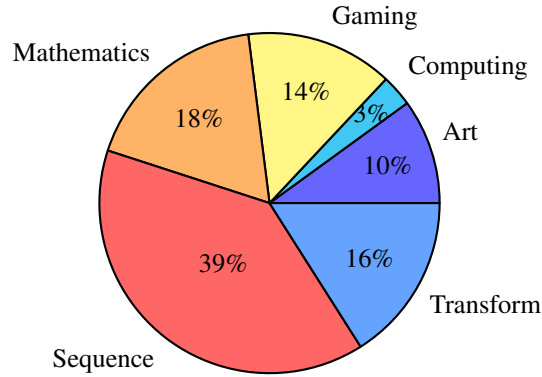
\begin{figure}[htbp]
    \centering
    \begin{tikzpicture}[scale=0.75]
    \pie{
        10/Art,
        3/Computing,
        14/Gaming,
        18/Mathematics,
        39/Sequence,
        16/Transform
    }
    \end{tikzpicture}
    \caption{Problem categories.}
    \label{fig:problem_categories}
  \end{figure}

The benchmark comprises of 6 primary categories:
\begin{itemize}
  \item Art: Text-based visual art challenges to test a model's ability to produce precise output formatting.
  \item Computing: Testing a model's understanding of computational principles.
  \item Gaming: Implementation of game rules and mechanics, Puzzle-solving challenges that require complex logical reasoning.
  \item Mathematics: Challenges focused on generating or manipulating mathematical sequences
  \item Sequence: Problems focused on text processing and string operations, requiring precise character-level manipulations.
  \item Transform: Problems requiring conversion between different representations.
\end{itemize}

\subsection{Solution Generation}

\begin{figure}
\centering
\begin{tabular}{|p{14cm}|}
\hline
\begin{minipage}{13.8cm}
You are an expert code golfer, tasked with solving programming challenges using the minimum 
number of characters or bytes possible while maintaining functionality. Your goal is to provide 
correct, extremely concise solutions to coding problems.

Here's the code golf challenge you need to solve:

problem\_name          : \{\{PROBLEM\_NAME\}\}

problem\_description   : 
\{\{PROBLEM\_DESCRIPTION\}\}

programming\_language  : \{\{LANGUAGE\}\}

\{\{EXAMPLE\}\}

Remember these key points:
- Ensure that the solution will execute within the specified time limit of 5 seconds.
- Ensure your final code is free of syntax errors.

Provide your solution in the following format:
Your ultra-concise final code should start with triple backquotes (\{\{LANGUAGE\}\}) and end 
in triple backquotes.
\end{minipage} \\
\hline
\end{tabular}
\caption{Prompt template.}
\label{fig:prompt_template}
\end{figure}
For solution generation, the prompt template shown in Figure~\ref{fig:prompt_template} is utilized. This template explicitly instructs the model to prioritize character minimization the primary objective in code golf while ensuring functional correctness. Each prompt includes the problem name, description, target programming language, and examples, providing comprehensive context for the model to generate optimally concise solutions. The primary goal in code golf. For each language, the implementation includes basic information about input/output handling specific to that language. For example, a Python solution needs to handle standard input differently than a Java solution.

\textbf{Generation Parameters}:\begin{itemize}
\item \textbf{Temperature}: 0.2
\item \textbf{Max tokens}: 32,768
\item \textbf{Top-p}: 0.95
\item \textbf{No of Samples Generated per problem}: 10
\end{itemize}

\subsection{Evaluation Methodology}
CodeGolf Bench employs a comprehensive evaluation methodology that includes solution validation, human baseline comparison and metrics calculation.
\subsubsection{Solution Validation}
Solutions are extracted from LLM responses and submitted to the \url{code.golf} API, which executes them against standardized test cases. The platform returns pass/fail status, error messages, and execution details. Only solutions that pass all test cases receive non-zero scores.
\subsubsection{Human Baseline Comparison}
Each solution is positioned within the distribution of human-written solutions from the \url{code.golf} platform:
\begin{enumerate}
\item Retrieve all existing human solutions via the API
\item Calculate distribution statistics (min/max/mean character counts)
\item Determine the model's percentile ranking in this distribution
\end{enumerate}
This approach normalizes for problem difficulty and language characteristics, enabling direct comparison between LLM and human performance.
\subsubsection{Metrics Calculation}
CodeGolf Bench calculates several metrics to evaluate solution quality:
\begin{itemize}
\item \textbf{Character Count}: Primary optimization target in code golf
\item \textbf{Byte Count}: Size when encoded in UTF-8, accounting for multi-byte characters
\item \textbf{Percentile Ranking}: Position within the distribution of human solutions, with 0\% indicating the model performs worse than all humans and 100\% indicating it performs better than all humans.
\item \textbf{Pass@k}: Success rate when considering the best of k attempts (calculated for k=1 through k=8)
\item \textbf{Global Percentile}: Average percentile across problems.
\end{itemize}
\subsubsection{Scoring System}
For each problem-language-model combination:
\begin{enumerate}
\item Solutions that fail to pass any test case receive a percentile score of 0.
\item For successful solutions, percentile is calculated as:
\begin{verbatim}
percentile = (position / total_solutions) * 100
\end{verbatim}
\item The best-scoring solution among the attempts is selected for the best-percentile score.
\item Pass@k metrics (k=1 to 8) are calculated.
\end{enumerate}
This scoring system rewards both correctness and efficiency, allowing fair comparison of models with different strengths. 

\section{Evaluation}

In this evaluation, the performance of nine large language models on Python and C++ coding tasks using the CodeGolf benchmark is assessed. The analysis focuses on performance across languages, success rates, and error distributions.

\subsection{Performance Differences Across Languages and Model Capabilities}

\begin{table}[htbp]
  \caption{Model Performance on Code Golf Tasks}
  \label{tab:model_performance}
  \centering
  \begin{tabular}{@{}lcccccc@{}}
  \toprule
  \multirow{2}{*}{Model} & \multicolumn{3}{c}{C++} & \multicolumn{3}{c}{Python} \\
  \cmidrule(lr){2-4} \cmidrule(lr){5-7}
  & Pass@1 & Pass@2 & Best \%ile
  & Pass@1 & Pass@2 & Best \%ile \\
  \midrule
  \multicolumn{7}{c}{\textbf{Reasoning Models}} \\
  \midrule
  Gemini-2.5-pro-preview-03-25 & 25.20 & 38.20 & 73.90
  & 38.20 & 53.04 & 68.04 \\
  DeepSeek-R1 & 39.13 & 48.69 & 60.21
  & 43.47 & 54.08 & 67.09 \\
  Qwen3-235B & 37.02 & 48.19 & 48.24
  & 32.94 & 43.33 & 41.64 \\
  DeepSeek-R1-\\Distill-Llama-70B & 23.41 & 31.71 & 44.18
  & 29.69 & 37.65 & 42.33 \\
  \midrule
  \multicolumn{7}{c}{\textbf{Non-Reasoning Models}} \\
  \midrule
  Gemini-2.5-flash-preview-04-17 & 43.24 & 54.21 & 55.19
  & 34.18 & 47.18 & 46.42 \\
  DeepSeek-V3-0324 & 19.46 & 24.77 & 38.00
  & 34.78 & 42.60 & 51.60 \\
  Llama-4-Maverick & 20 & 27.82 & 44.85
  & 38.69 & 44.21 & 55.31 \\
  Llama-4-Scout & 6 & 8.6 & 16.65
  & 30.34 & 34.79 & 36.69 \\
  Gemma-3-27b-it & 24.49 & 24.93 & 20.92
  & 23.47 & 26.95 & 25.23 \\
  \bottomrule
  \end{tabular}
\end{table}

Table~\ref{tab:model_performance} presents Pass@1 and Pass@2 metrics alongside percentile rankings relative to human solutions. The performance metrics show notable differences between model categories. Reasoning models generally achieve higher percentile scores than other models, particularly in C++.

Reasoning models achieve higher Best Percentile scores in C++ (73.90\% and 60.21\%) compared to the highest-scoring non-reasoning model, Gemini-2.5-flash-preview-04-17 (55.19\%). The performance gap between reasoning and non-reasoning models is more pronounced in C++ than in Python. In C++, Gemma-3-27B performs considerably worse, with scores as low as 20.92\% compared to Gemini-2.5-pro-preview-03-25 at 73.90\%. The performance gap between the highest and lowest performing models is substantially larger in C++ (57.25 percentage points) than in Python (42.81 percentage points).

The improvement between Pass@1 and Pass@2 varies significantly across models. In C++, Gemini-2.5-flash-preview-04-17 and Qwen3-235B show the largest absolute improvement (+10.97 and +11.17 percentage points), while Gemma-3-27b-it shows minimal change (+0.44 points), DeepSeek-R1-Distill-Llama-70B model shows a more modest gain (+8.3 points). In Python, Gemini-2.5-pro-preview-03-25 demonstrates the largest improvement (+14.84 points), with Qwen3-235B showing a similar improvement (+10.39 points).

Language-specific performance differences are notable, with some models performing better in Python (DeepSeek-V3-0324, Llama-4-Maverick), while others show stronger results in C++ (Gemini-2.5-flash-preview-04-17, Qwen3-235B). In Python, the performance spread is more moderate. DeepSeek-R1 achieves 67.09\% while Gemini-2.5-flash-preview-04-17 reaches a percentile score of 46.42\%. The data suggests that Python's more accessible syntax and dynamic typing may reduce the performance gap between different model categories compared to C++.

Further analysis of Pass@3 through Pass@8 metrics (shown in Tables~\ref{tab:cpp_performance} and \ref{tab:python_performance}) reveals interesting patterns. In C++, DeepSeek-R1 consistently outperforms other models, with Pass@8 reaching 66.55\%. Similarly, in Python tasks, Gemini-2.5-pro-preview-03-25 demonstrates strong performance, with comparable or slightly better metrics than DeepSeek-R1 across most Pass@k measurements.

\begin{figure}[htbp]
    \centering
  \begin{tikzpicture}[scale=0.75]
    \centering
    \begin{axis}[
        ybar,
        width=\linewidth,
        height=7cm,
        bar width=12pt,
        ylabel={Percentile},
        ymin=0,
        ymax=100,
        enlarge y limits={upper=0.05},
        % Numerical X-Axis Setup
        xtick={1,...,9},
        xticklabels={
            Gemini-2.5-pro-preview-03-25, DeepSeek-R1, Gemini-2.5-flash-preview-04-17, Llama-4-Maverick, Qwen3-235B, DeepSeek-V3-0324, DeepSeek-R1-Distill-Llama-70B, Llama-4-Scout, Gemma-3-27b-it
        },
        % Modified label styling to be near-vertical and more compact
        xticklabel style={
            rotate=30,           % Fully vertical labels
            anchor=east,         % Anchor at the east (right) side
            font=\scriptsize,    % Small font size
            inner sep=1pt,       % Reduced inner padding
        },
        % Explicitly disable any bar shift to ensure ticks align with bar centers
        bar shift=0pt,
        enlarge x limits=0.05,
        nodes near coords,
        nodes near coords align={vertical},
        nodes near coords style={font=\tiny, rotate=0, anchor=south},
        point meta=y,
        ymajorgrids=true,
        grid style={dashed, gray!50},
        % --- Legend Styling ---
        legend style={
            at={(0.97,0.97)},     % Position of the legend in top right
            anchor=north east,    % Anchor point for positioning
            legend columns=1,     % Single column (vertical) layout
            draw=black,           % Add border
            fill=white,           % White background
            row sep=2pt,          % Space between rows
            legend cell align=left, % Left align text
        },
        legend image code/.code={
            \draw[#1, draw=black] (0pt,-1.5mm) rectangle (5mm,1.5mm);
        },
    ]
  
  % Gemini-2.5-pro-preview-03-25-preview-03-25
  \pgfmathsetmacro{\percentilePythonGemini-2.5-Pro-preview-03-25}{68.04130434782608}
  \pgfmathsetmacro{\percentileCPPGemini-2.5-Pro-preview-03-25}{73.90660869565217}
  % (68.04130434782608 + 73.90660869565217) / 2 = 70.97395652173913
  
  % DeepSeek-R1
  \pgfmathsetmacro{\percentilePythonDeepSeek-R1}{67.09947826086956}
  \pgfmathsetmacro{\percentileCPPDeepSeek-R1}{60.21791304347826}
  % (67.09947826086956 + 60.21791304347826) / 2 = 63.65869565217391
  
  % Qwen3-235B
  \pgfmathsetmacro{\percentilePythonQwen}{41.636339285714286}
  \pgfmathsetmacro{\percentileCPPQwen}{48.238727272727274}
  % (41.636339285714286 + 48.238727272727274) / 2 = 44.93753328

  % DeepSeek-R1-Distill-Llama-70B
  \pgfmathsetmacro{\percentilePythonDeepSeekDistill}{42.32791304347826}
  \pgfmathsetmacro{\percentileCPPDeepSeekDistill}{44.179391304347824}
  % (42.32791304347826 + 44.179391304347824) / 2 = 43.25365217391304
  
  % gemini-2.5-flash-preview-04-17
  \pgfmathsetmacro{\percentilePythonGemini-2.5-flash-preview-04-17}{46.423762376237626}
  \pgfmathsetmacro{\percentileCPPGemini-2.5-flash-preview-04-17}{55.19149122807018}
  % (46.423762376237626 + 55.19149122807018) / 2 = 50.807626802153904
  
  % Llama-4-Maverick
  \pgfmathsetmacro{\percentilePythonLlama-4-Maverick}{55.31}
  \pgfmathsetmacro{\percentileCPPLlama-4-Maverick}{44.854608695652175}
  % (55.31 + 44.854608695652175) / 2 = 50.08230434782608
  
  % DeepSeek-V3-0324
  \pgfmathsetmacro{\percentilePythonDeepSeek-V3-0324}{51.608695652173914}
  \pgfmathsetmacro{\percentileCPPDeepSeek-V3-0324}{38.0058407079646}
  % (51.608695652173914 + 38.0058407079646) / 2 = 44.80726818006926
  
  % Llama-4-Scout
  \pgfmathsetmacro{\percentilePythonLlama-4-Scout}{36.695913043478264}
  \pgfmathsetmacro{\percentileCPPLlama-4-Scout}{16.652347826086956}
  % (36.695913043478264 + 16.652347826086956) / 2 = 26.67413043478261
  
  % Gemma-3-27b-it
  \pgfmathsetmacro{\percentilePythonGemma-3-27b-it}{25.231739130434782}
  \pgfmathsetmacro{\percentileCPPGemma-3-27b-it}{20.92069565217391}
  % (25.231739130434782 + 20.92069565217391) / 2 = 23.076217391304346
  
  % Bar Data
  \addplot+[ybar, fill=blue!40, draw=black] coordinates {
      (1, 70.97395652173913)  % Gemini-2.5-pro-preview-03-25-preview-03-25
      (2, 63.65869565217391)  % DeepSeek-R1
      (5, 44.93753328)  % Qwen3-235B
      (7, 43.25365217391304)  % DeepSeek-R1-Distill-Llama-70B
      };
  \addlegendentry{Reasoning}
  \addplot+[ybar, fill=red!40, draw=black] coordinates {
      (3, 50.807626802153904)  % gemini-2.5-flash-preview-04-17
      (4, 50.08230434782608)  % Llama-4-Maverick
      (6, 44.80726818006926)  % DeepSeek-V3-0324
      (8, 26.67413043478261)  % Llama-4-Scout
      (9, 23.076217391304346)  % Gemma-3-27b-it
  };
  \addlegendentry{Non-Reasoning}
  
  \end{axis}
  \end{tikzpicture}
  \caption{Average of best C++ and Python percentiles}
  \label{fig:barchart}
\end{figure}
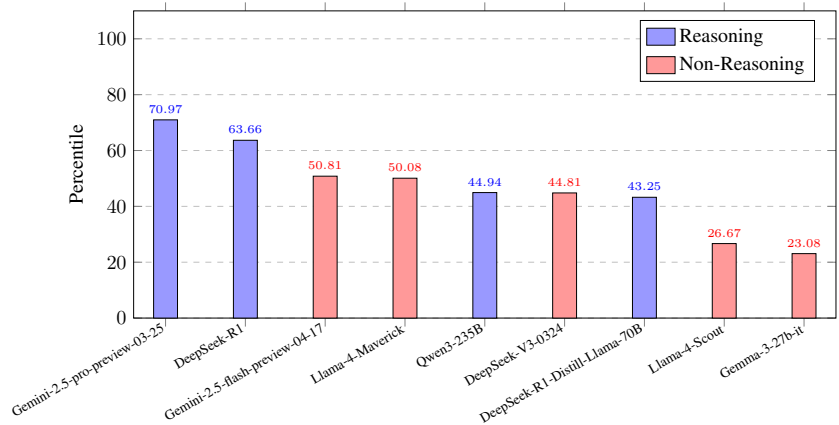

Figure~\ref{fig:barchart} visualizes the average best percentile scores across both languages, providing a clear comparative view of overall model performance.

\subsection{Category-Specific Performance}

\begin{figure}[htbp]
  \centering
  \begin{subfigure}{0.48\textwidth}
  \centering
  \begin{tikzpicture}[scale=0.8]
  \begin{polaraxis}[
    width=7cm,
    height=7cm,
    xtick={0.0,60.0,120.0,180.0,240.0,300.0},
    xticklabels={\textbf{Art},\textbf{Computing},\textbf{Gaming},\textbf{Mathematics},\textbf{Sequence},\textbf{Transform}},
    xticklabel style={font=\tiny, anchor=center, inner sep=0.5mm, outer sep=0.5mm},
    ytick={0,20,40,60,80,100},
    yticklabels={0\%,20\%,40\%,60\%,80\%,100\%},
    yticklabel style={font=\tiny},
    ymin=0,
    ymax=100,
    grid=both,
    grid style={line width=.05pt, draw=gray!10},
    major grid style={line width=.1pt,draw=gray!50},
  ]
  \addplot[color=blue,mark=*,line width=0.5pt,solid] coordinates {(0.0,28.82) (60.0,46.41) (120.0,21.43) (180.0,29.87) (240.0,54.87) (300.0,27.60)};
  \addplot[color=cyan,mark=square*,line width=0.5pt,solid] coordinates {(0.0,70.00) (60.0,55.92) (120.0,26.80) (180.0,42.34) (240.0,91.38) (300.0,41.40)};
  \addplot[color=orange,mark=triangle*,line width=0.5pt,solid] coordinates {(0.0,62.54) (60.0,46.70) (120.0,30.29) (180.0,47.96) (240.0,76.86) (300.0,40.62)};
  \addplot[color=red,mark=diamond*,line width=0.5pt,solid] coordinates {(0.0,60.00) (60.0,75.00) (120.0,70.71) (180.0,80.20) (240.0,89.21) (300.0,53.33)};
  \addplot[color=brown,mark=triangle*,line width=0.5pt,dashdotted] coordinates {(0.0,57.04) (60.0,46.40) (120.0,18.52) (180.0,31.63) (240.0,70.90) (300.0,35.90)};
  \addplot[color=teal,mark=triangle*,line width=0.5pt,dashdotted] coordinates {(0.0,37.16) (60.0,23.23) (120.0,19.26) (180.0,25.56) (240.0,76.96) (300.0,30.47)};
  \addplot[color=purple,mark=o,line width=0.5pt,dashed] coordinates {(0.0,10.00) (60.0,37.50) (120.0,14.29) (180.0,27.78) (240.0,78.42) (300.0,38.46)};
  \addplot[color=violet,mark=pentagon*,line width=0.5pt,dashed] coordinates {(0.0,18.91) (60.0,0.00) (120.0,0.00) (180.0,5.56) (240.0,39.13) (300.0,3.85)};
  \addplot[color=green,mark=x,line width=0.5pt,dashdotted] coordinates {(0.0,10.00) (60.0,0.00) (120.0,0.00) (180.0,0.00) (240.0,54.17) (300.0,7.44)};
  \end{polaraxis}
  \end{tikzpicture}
  \caption{Best Percentile (C++)}
  \end{subfigure}
  \hfill
  \begin{subfigure}{0.48\textwidth}
  \centering
  \begin{tikzpicture}[scale=0.75]
  \begin{polaraxis}[
    width=7cm,
    height=7cm,
    xtick={0.0,60.0,120.0,180.0,240.0,300.0},
    xticklabels={\textbf{Art},\textbf{Computing},\textbf{Gaming},\textbf{Mathematics},\textbf{Sequence},\textbf{Transform}},
    xticklabel style={font=\tiny, anchor=center, inner sep=0.5mm, outer sep=0.5mm},
    ytick={0,20,40,60,80,100},
    yticklabels={0\%,20\%,40\%,60\%,80\%,100\%},
    yticklabel style={font=\tiny},
    ymin=0,
    ymax=100,
    grid=both,
    grid style={line width=.05pt, draw=gray!10},
    major grid style={line width=.1pt,draw=gray!50},
  ]
  \addplot[color=blue,mark=*,line width=0.5pt,solid] coordinates {(0.0,20.00) (60.0,35.00) (120.0,17.86) (180.0,50.00) (240.0,83.76) (300.0,39.94)};
  \addplot[color=cyan,mark=square*,line width=0.5pt,solid] coordinates {(0.0,80.00) (60.0,59.43) (120.0,40.22) (180.0,44.44) (240.0,94.31) (300.0,53.85)};
  \addplot[color=orange,mark=triangle*,line width=0.5pt,solid] coordinates {(0.0,50.88) (60.0,37.64) (120.0,31.31) (180.0,43.64) (240.0,63.52) (300.0,30.65)};
  \addplot[color=red,mark=diamond*,line width=0.5pt,solid] coordinates {(0.0,80.00) (60.0,61.36) (120.0,42.86) (180.0,55.56) (240.0,96.67) (300.0,44.76)};
  \addplot[color=brown,mark=triangle*,line width=0.5pt,dashdotted] coordinates {(0.0,53.01) (60.0,33.44) (120.0,20.82) (180.0,41.77) (240.0,52.60) (300.0,34.18)};
  \addplot[color=teal,mark=triangle*,line width=0.5pt,dashdotted] coordinates {(0.0,35.49) (60.0,30.34) (120.0,15.34) (180.0,34.68) (240.0,65.01) (300.0,34.45)};
  \addplot[color=purple,mark=o,line width=0.5pt,dashed] coordinates {(0.0,26.03) (60.0,37.11) (120.0,41.17) (180.0,50.04) (240.0,78.48) (300.0,48.68)};
  \addplot[color=violet,mark=pentagon*,line width=0.5pt,dashed] coordinates {(0.0,16.10) (60.0,24.73) (120.0,0.00) (180.0,24.24) (240.0,68.13) (300.0,29.53)};
  \addplot[color=green,mark=x,line width=0.5pt,dashdotted] coordinates {(0.0,9.54) (60.0,36.68) (120.0,6.68) (180.0,0.00) (240.0,52.69) (300.0,14.02)};
  \end{polaraxis}
  \end{tikzpicture}
  \caption{Best Percentile (Python)}
  \end{subfigure}
  
  \vspace{0.5em}
  \centering
  \begin{tabular}{@{}clclcl@{}}
    \textcolor{blue}{$\bullet$} & DeepSeek-V3-0324 &
    \textcolor{cyan}{$\blacksquare$} & DeepSeek-R1 &
    \textcolor{orange}{$\blacktriangle$} & Gemini-2.5-flash \\
    \textcolor{red}{$\blacklozenge$} & Gemini-2.5-pro-preview-03-25 &
    \textcolor{brown}{$\blacktriangledown$} & Qwen3-235B &
    \textcolor{teal}{$\blacktriangleleft$} & DeepSeek-R1-Distill-Llama-70B \\
    \textcolor{purple}{$\bullet$} & Llama-4-Maverick &
    \textcolor{violet}{$\bigstar$} & Llama-4-Scout &
    \textcolor{green}{$\times$} & Gemma-3-27b \\
\end{tabular}
  
  \caption{Best Percentile Scores by Category for Different Models}
  \label{fig:radar_charts}
\end{figure}
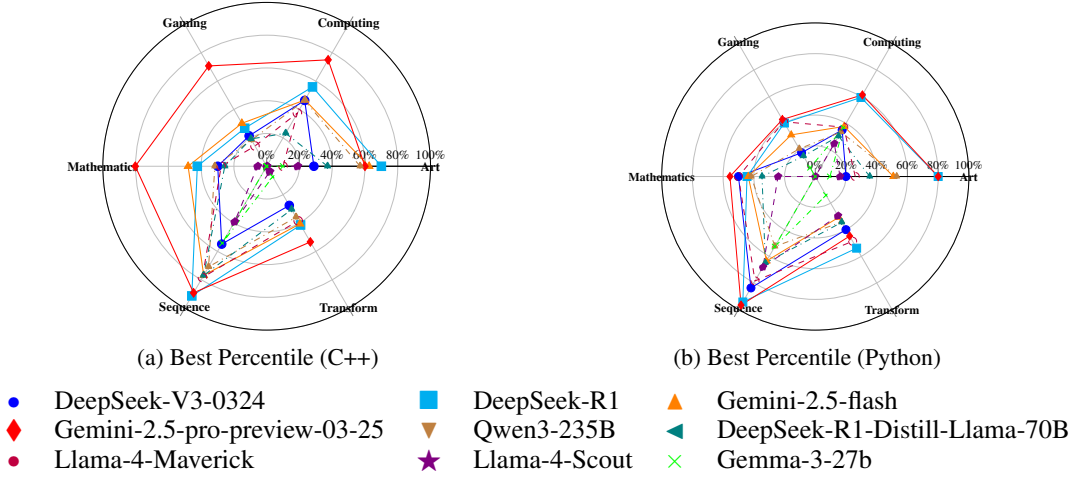

Figure~\ref{fig:radar_charts} reveals significant performance differences between reasoning and non-reasoning models across task categories. For both C++ and Python, Sequence tasks generally yield the highest scores, but with notable disparities between model types.

In C++, reasoning models show superior performance: Gemini-2.5-pro-preview-03-25 and DeepSeek-R1 excel in Sequence (89-91\%) and Mathematics (80-42\%), while Qwen3-235B and DeepSeek-R1-Distill-Llama-70B achieve moderate scores (71-77\%) in Sequence. Non-reasoning models exhibit greater variability, with Llama-4-Maverick reaching 78\% in Sequence but only 14\% in Gaming, and both Gemma-3-27B and Llama-4-Scout scoring 0\% in multiple categories.

In Python, reasoning models demonstrate more balanced performance across categories, with Gemini-2.5-pro-preview-03-25 and DeepSeek-R1 achieving their highest scores in Sequence (97\% and 94\% respectively) and Art (80\%). Non-reasoning models show stronger performance in Python's Sequence category (53-83\%) compared to other categories, with Llama-4-Maverick performing better than other non-reasoning models across most categories.

Consistent weaknesses are evident in non-reasoning models, particularly in Mathematics and Gaming. Gemma-3-27B scores 0\% in Mathematics for both languages and in Gaming for C++. Llama-4-Scout shows similar deficiencies with 0\% in Gaming for C++. Overall, reasoning models consistently outperform non-reasoning models across nearly all categories, with particularly pronounced gaps in Mathematics and Computing, suggesting that reasoning capabilities provide significant advantages for more complex programming tasks.

\subsection{Insights from Error Distributions}

\begin{figure}[htbp]
  \centering
  \begin{subfigure}{0.498\textwidth} % Adjusted width for smaller gap
    \centering
    \begin{tikzpicture}[scale=0.5]
    \begin{axis}[
        xbar stacked,
        xlabel={Error Distribution (\% of Total Errors)},
        symbolic y coords={DeepSeek-R1-Py,Gemini-2.5-pro-preview-03-25-Py,Llama-4-Maverick-17B-Py,Gemini-2.5-flash-preview-04-17-Py,Qwen3-235B-Py,DeepSeek-R1-Distill-Llama-70B-Py,Deepseek-V3-0324-Py,Llama-4-Scout-17B-Py,Gemma-3-27B-Py},
        ytick=data,
        yticklabels={\textbf{DeepSeek-R1},\textbf{Gemini-2.5-pro-preview-03-25},\textbf{Llama-4-Maverick-17B},\textbf{Gemini-2.5-flash-preview-04-17},\textbf{Qwen3-235B},\textbf{DeepSeek-R1-Distill-Llama-70B},\textbf{Deepseek-V3-0324},\textbf{Llama-4-Scout-17B},\textbf{Gemma-3-27B}},
        legend style={legend columns=6, font=\tiny, legend pos=south west, anchor=north west, at={(0.05,-0.25)}}, % Changed legend columns to 6 for single row
        legend to name=globallegend, % Name the legend to reference it later
        yticklabel style={font=\footnotesize, rotate=45, anchor=east}, % Made labels slanted and bigger
        xmin=0,
        xmax=100,
        bar width=5pt,
        width=9cm,
        height=7cm,
        y dir=reverse,
        nodes near coords align={horizontal},
        every node near coord/.append style={font=\tiny},
    ]
    
    % AssertionError for all models - normalized to % of errors
    \addplot[fill=blue] coordinates 
        {(0.0,DeepSeek-R1-Py)
         (0.0,Gemini-2.5-pro-preview-03-25-Py)
         (0.0,Llama-4-Maverick-17B-Py)
         (0.0,Gemini-2.5-flash-preview-04-17-Py)
         (0.0,Qwen3-235B-Py)
         (0.0,DeepSeek-R1-Distill-Llama-70B-Py)
         (0.0,Deepseek-V3-0324-Py)
         (0.0,Llama-4-Scout-17B-Py)
         (0.0,Gemma-3-27B-Py)};
    \addlegendentry{AssertionError}
    
    % NameError for all models - normalized to % of errors
    \addplot[fill=orange] coordinates 
        {(0.9,DeepSeek-R1-Py)
         (1.3,Gemini-2.5-pro-preview-03-25-Py)
         (7.2,Llama-4-Maverick-17B-Py)
         (1.6,Gemini-2.5-flash-preview-04-17-Py)
         (1.7,Qwen3-235B-Py)
         (1.5,DeepSeek-R1-Distill-Llama-70B-Py)
         (4.6,Deepseek-V3-0324-Py)
         (12.5,Llama-4-Scout-17B-Py)
         (4.5,Gemma-3-27B-Py)};
    \addlegendentry{NameError}
    
    % ValueError for all models - normalized to % of errors
    \addplot[fill=green!60!black] coordinates 
        {(10.5,DeepSeek-R1-Py)
         (10.1,Gemini-2.5-pro-preview-03-25-Py)
         (13.2,Llama-4-Maverick-17B-Py)
         (3.9,Gemini-2.5-flash-preview-04-17-Py)
         (7.4,Qwen3-235B-Py)
         (4.1,DeepSeek-R1-Distill-Llama-70B-Py)
         (18.9,Deepseek-V3-0324-Py)
         (4.0,Llama-4-Scout-17B-Py)
         (12.3,Gemma-3-27B-Py)};
    \addlegendentry{ValueError}
    
    % IndexError for all models - normalized to % of errors
    \addplot[fill=red] coordinates 
        {(9.3,DeepSeek-R1-Py)
         (7.7,Gemini-2.5-pro-preview-03-25-Py)
         (17.2,Llama-4-Maverick-17B-Py)
         (11.7,Gemini-2.5-flash-preview-04-17-Py)
         (5.6,Qwen3-235B-Py)
         (1.9,DeepSeek-R1-Distill-Llama-70B-Py)
         (16.2,Deepseek-V3-0324-Py)
         (5.7,Llama-4-Scout-17B-Py)
         (8.7,Gemma-3-27B-Py)};
    \addlegendentry{IndexError}
    
    % TypeError for all models - normalized to % of errors
    \addplot[fill=purple] coordinates 
        {(5.6,DeepSeek-R1-Py)
         (1.6,Gemini-2.5-pro-preview-03-25-Py)
         (7.2,Llama-4-Maverick-17B-Py)
         (1.8,Gemini-2.5-flash-preview-04-17-Py)
         (3.2,Qwen3-235B-Py)
         (2.3,DeepSeek-R1-Distill-Llama-70B-Py)
         (8.9,Deepseek-V3-0324-Py)
         (4.4,Llama-4-Scout-17B-Py)
         (5.5,Gemma-3-27B-Py)};
    \addlegendentry{TypeError}
    
    % OtherError for all models - normalized to % of errors
    \addplot[fill=brown!70!black] coordinates 
        {(73.7,DeepSeek-R1-Py)
         (79.3,Gemini-2.5-pro-preview-03-25-Py)
         (55.2,Llama-4-Maverick-17B-Py)
         (81.0,Gemini-2.5-flash-preview-04-17-Py)
         (82.1,Qwen3-235B-Py)
         (90.2,DeepSeek-R1-Distill-Llama-70B-Py)
         (51.4,Deepseek-V3-0324-Py)
         (73.4,Llama-4-Scout-17B-Py)
         (69.0,Gemma-3-27B-Py)};
    \addlegendentry{OtherError}
    
    \end{axis}
    \end{tikzpicture}
    \caption{Python}
  \end{subfigure}\hfill%
  \begin{subfigure}{0.498\textwidth} % Adjusted width for smaller gap
    \centering
    \begin{tikzpicture}[scale=0.5]
    \begin{axis}[
        xbar stacked,
        xlabel={Error Distribution (\% of Total Errors)},
        symbolic y coords={Gemini-2.5-flash-preview-04-17-CPP,DeepSeek-R1-CPP,Qwen3-235B-CPP,Gemini-2.5-pro-preview-03-25-CPP,DeepSeek-R1-Distill-Llama-70B-CPP,Gemma-3-27B-CPP,Llama-4-Maverick-17B-CPP,Deepseek-V3-0324-CPP,Llama-4-Scout-17B-CPP},
        ytick=data, % Keep ytick=data for symbolic y coords alignment
        yticklabels={\textbf{DeepSeek-R1},\textbf{Gemini-2.5-pro-preview-03-25},\textbf{Llama-4-Maverick-17B},\textbf{Gemini-2.5-flash-preview-04-17},\textbf{Qwen3-235B},\textbf{DeepSeek-R1-Distill-Llama-70B},\textbf{Deepseek-V3-0324},\textbf{Llama-4-Scout-17B},\textbf{Gemma-3-27B}}, % Remove y-tick labels
        yticklabel style={font=\footnotesize, rotate=45, anchor=east}, % Made labels slanted and bigger
        xmin=0,
        xmax=100,
        bar width=5pt,
        width=9cm,
        height=7cm,
        % yticklabel style removed as labels are gone
        y dir=reverse,
        nodes near coords align={horizontal},
        every node near coord/.append style={font=\tiny},
    ]
    
    % AssertionError for all models - normalized to % of errors
    \addplot[fill=blue] coordinates 
        {(0.0,Gemini-2.5-flash-preview-04-17-CPP)
         (0.0,DeepSeek-R1-CPP)
         (0.1,Qwen3-235B-CPP)
         (0.2,Gemini-2.5-pro-preview-03-25-CPP)
         (0.0,DeepSeek-R1-Distill-Llama-70B-CPP)
         (0.0,Gemma-3-27B-CPP)
         (0.0,Llama-4-Maverick-17B-CPP)
         (0.0,Deepseek-V3-0324-CPP)
         (0.0,Llama-4-Scout-17B-CPP)};
    % No \addlegendentry here
    
    % NameError for all models - normalized to % of errors
    \addplot[fill=orange] coordinates 
        {(14.5,Gemini-2.5-flash-preview-04-17-CPP)
         (8.6,DeepSeek-R1-CPP)
         (22.2,Qwen3-235B-CPP)
         (13.4,Gemini-2.5-pro-preview-03-25-CPP)
         (13.2,DeepSeek-R1-Distill-Llama-70B-CPP)
         (21.4,Gemma-3-27B-CPP)
         (23.6,Llama-4-Maverick-17B-CPP)
         (6.6,Deepseek-V3-0324-CPP)
         (4.0,Llama-4-Scout-17B-CPP)};
    % No \addlegendentry here
    
    % ValueError for all models - normalized to % of errors
    \addplot[fill=green!60!black] coordinates 
        {(2.2,Gemini-2.5-flash-preview-04-17-CPP)
         (4.3,DeepSeek-R1-CPP)
         (2.5,Qwen3-235B-CPP)
         (0.0,Gemini-2.5-pro-preview-03-25-CPP)
         (2.3,DeepSeek-R1-Distill-Llama-70B-CPP)
         (0.5,Gemma-3-27B-CPP)
         (2.9,Llama-4-Maverick-17B-CPP)
         (1.5,Deepseek-V3-0324-CPP)
         (0.2,Llama-4-Scout-17B-CPP)};
    % No \addlegendentry here
    
    % IndexError for all models - normalized to % of errors
    \addplot[fill=red] coordinates 
        {(2.2,Gemini-2.5-flash-preview-04-17-CPP)
         (0.0,DeepSeek-R1-CPP)
         (0.2,Qwen3-235B-CPP)
         (0.2,Gemini-2.5-pro-preview-03-25-CPP)
         (0.0,DeepSeek-R1-Distill-Llama-70B-CPP)
         (0.0,Gemma-3-27B-CPP)
         (0.0,Llama-4-Maverick-17B-CPP)
         (0.0,Deepseek-V3-0324-CPP)
         (0.0,Llama-4-Scout-17B-CPP)};
    % No \addlegendentry here
    
    % TypeError for all models - normalized to % of errors
    \addplot[fill=purple] coordinates 
        {(3.5,Gemini-2.5-flash-preview-04-17-CPP)
         (5.6,DeepSeek-R1-CPP)
         (3.7,Qwen3-235B-CPP)
         (2.7,Gemini-2.5-pro-preview-03-25-CPP)
         (2.5,DeepSeek-R1-Distill-Llama-70B-CPP)
         (0.9,Gemma-3-27B-CPP)
         (3.3,Llama-4-Maverick-17B-CPP)
         (3.1,Deepseek-V3-0324-CPP)
         (0.2,Llama-4-Scout-17B-CPP)};
    % No \addlegendentry here
    
    % OtherError for all models - normalized to % of errors
    \addplot[fill=brown!70!black] coordinates 
        {(77.6,Gemini-2.5-flash-preview-04-17-CPP)
         (81.5,DeepSeek-R1-CPP)
         (71.3,Qwen3-235B-CPP)
         (83.5,Gemini-2.5-pro-preview-03-25-CPP)
         (82.0,DeepSeek-R1-Distill-Llama-70B-CPP)
         (77.2,Gemma-3-27B-CPP)
         (70.2,Llama-4-Maverick-17B-CPP)
         (88.8,Deepseek-V3-0324-CPP)
         (95.6,Llama-4-Scout-17B-CPP)};
    % No \addlegendentry here
    
    \end{axis}
    \end{tikzpicture}
    \caption{C++}
  \end{subfigure}
  
  \centering % Center the legend
  \ref{globallegend} % Place the shared legend

  \caption{Distribution of Error Types by Model (as percentage of total errors)} % Unnumbered caption
  \label{fig:error_distributions} % Label can be kept for cross-referencing if needed
\end{figure}
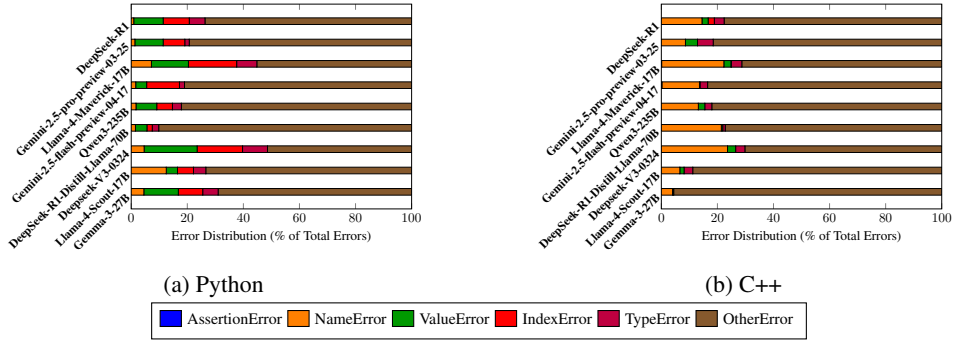

Figure~\ref{fig:error_distributions} shows the error type distribution across Python and C++. "OtherError" dominates, ranging from 51.4\% to 95.6\%, with DeepSeek-R1-Distill-Llama-70B exhibiting a notably high rate in Python (90.2\%) and Qwen3-235B showing a high NameError rate in C++ (22.2\%).

C++ implementations typically have higher NameError rates (e.g., Llama-4-Maverick: 23.6\% vs. 7.2\% in Python), while Python implementations show elevated ValueError and IndexError rates (e.g., DeepSeek-V3-0324: 18.9\% ValueError in Python vs. 1.5\% in C++). Among Python models, DeepSeek-R1 and Gemini-2.5-pro-preview-03-25 have the lowest NameError rates (0.9\% and 1.3\%, respectively), suggesting robust variable management.
\subsection{Consistency Across Languages}

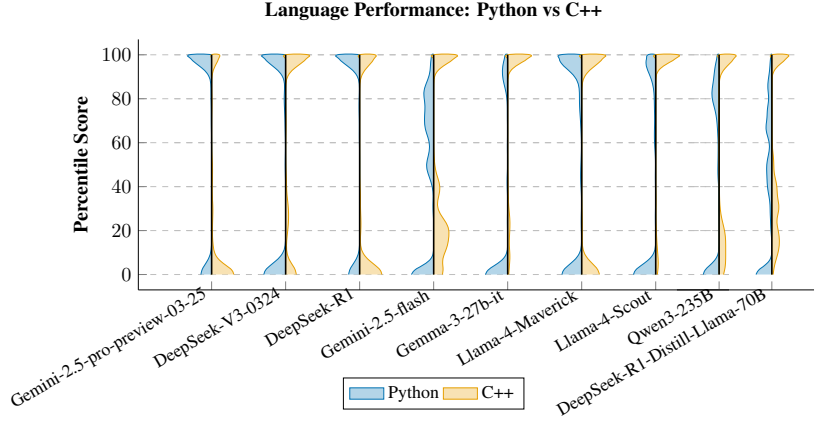
\begin{figure}[htbp]
  \centering
  \begin{tikzpicture}[scale=0.75]
    \begin{scope}[xshift=-1.5cm]
    \begin{axis}[
        name=plot1,
        title={Language Performance: Python vs C++},
        title style={font=\bfseries, align=left},
        axis y line*=left,
        axis x line*=bottom,
        width=12cm, height=6cm,
        scale only axis=false,
        enlarge x limits=0.07,
        enlarge y limits=0.07,
        xtick={0,1,2,3,4,5,6},
        xticklabels={
            {Gemini-2.5-pro-preview-03-25}, 
            {DeepSeek-V3-0324}, 
            {DeepSeek-R1}, 
            {Gemini-2.5-flash}, 
            {Gemma-3-27b-it}, 
            {Llama-4-Maverick}, 
            {Llama-4-Scout}
        },
        xticklabel style={
            rotate=30,
            anchor=east,
            font=\small,
        },
        xmin=-0.5, xmax=6.5,
        ytick={0,20,40,60,80,100},
        ymin=0, ymax=100,
        ylabel={Percentile Score},
        ylabel style={font=\bfseries},
        ymajorgrids=true,
        grid style={dashed, gray!50, line width=0.5pt},
        legend style={at={(0.5,-0.35)}, anchor=north, legend columns=2, draw=black, fill=white, fill opacity=1.0, text opacity=1, font=\small},
    ]
    
    % Adding explicit horizontal grid lines
    \draw[dashed, gray!50, line width=0.5pt] (axis cs:-0.5,20) -- (axis cs:6.5,20);
    \draw[dashed, gray!50, line width=0.5pt] (axis cs:-0.5,40) -- (axis cs:6.5,40);
    \draw[dashed, gray!50, line width=0.5pt] (axis cs:-0.5,60) -- (axis cs:6.5,60);
    \draw[dashed, gray!50, line width=0.5pt] (axis cs:-0.5,80) -- (axis cs:6.5,80);
    \draw[dashed, gray!50, line width=0.5pt] (axis cs:-0.5,100) -- (axis cs:6.5,100);
    
    % Gemini 2.5 Pro
    \addplot[python_blue, fill=python_blue!30, draw=python_blue, area legend,
              smooth, tension=0.8] table [x=x, y=y] {\pythongeminitwodata};
    \addlegendentry{Python}
    
    \addplot[cpp_orange, fill=cpp_orange!30, draw=cpp_orange, area legend,
              smooth, tension=0.8] table [x=x, y=y] {\cppgeminitwodata};
    \addlegendentry{C++}
    
    % DeepSeek-V3
    \addplot[python_blue, fill=python_blue!30, draw=python_blue, forget plot,
              smooth, tension=0.8] table [x=x, y=y] {\pythondeepseekdata};
    
    \addplot[cpp_orange, fill=cpp_orange!30, draw=cpp_orange, forget plot,
              smooth, tension=0.8] table [x=x, y=y] {\cppdeepseekdata};
    
    % DeepSeek Reasoner
    \addplot[python_blue, fill=python_blue!30, draw=python_blue, forget plot,
              smooth, tension=0.8] table [x=x, y=y] {\pythondeepseekreasonordata};
    
    \addplot[cpp_orange, fill=cpp_orange!30, draw=cpp_orange, forget plot,
              smooth, tension=0.8] table [x=x, y=y] {\cppdeepseekreasonordata};
    
    % Gemini 2.5 Flash
    \addplot[python_blue, fill=python_blue!30, draw=python_blue, forget plot,
              smooth, tension=0.8] table [x=x, y=y] {\pythongeminiflashdata};
    
    \addplot[cpp_orange, fill=cpp_orange!30, draw=cpp_orange, forget plot,
              smooth, tension=0.8] table [x=x, y=y] {\cppgeminiflashdata};
    
    % Gemma
    \addplot[python_blue, fill=python_blue!30, draw=python_blue, forget plot,
              smooth, tension=0.8] table [x=x, y=y] {\pythongemmarouter};
    
    \addplot[cpp_orange, fill=cpp_orange!30, draw=cpp_orange, forget plot,
              smooth, tension=0.8] table [x=x, y=y] {\cppgemmarouter};
    
    % Llama Maverick
    \addplot[python_blue, fill=python_blue!30, draw=python_blue, forget plot,
              smooth, tension=0.8] table [x=x, y=y] {\pythonllamamaverick};
    
    \addplot[cpp_orange, fill=cpp_orange!30, draw=cpp_orange, forget plot,
              smooth, tension=0.8] table [x=x, y=y] {\cppllamamaverick};
    
    % Llama Scout
    \addplot[python_blue, fill=python_blue!30, draw=python_blue, forget plot,
              smooth, tension=0.8] table [x=x, y=y] {\pythonllamascout};
    
    \addplot[cpp_orange, fill=cpp_orange!30, draw=cpp_orange, forget plot,
              smooth, tension=0.8] table [x=x, y=y] {\cppllamascout};
    
    % Draw model separators
    \addplot[black, thick] coordinates {(0, 0) (0, 100)};
    \addplot[black, thick] coordinates {(1, 0) (1, 100)};
    \addplot[black, thick] coordinates {(2, 0) (2, 100)};
    \addplot[black, thick] coordinates {(3, 0) (3, 100)};
    \addplot[black, thick] coordinates {(4, 0) (4, 100)};
    \addplot[black, thick] coordinates {(5, 0) (5, 100)};
    \addplot[black, thick] coordinates {(6, 0) (6, 100)};
    
    \end{axis}
    \end{scope}
    
    \begin{scope}[xshift=8cm]
    \begin{axis}[
        name=plot2,
        title={},
        axis y line=none,
        axis x line*=bottom,
        width=4cm, height=6cm,
        scale only axis=false,
        enlarge x limits=0.15,
        enlarge y limits=0.07,
        xtick={0,1},
        xticklabels={
            {Qwen3-235B}, 
            {DeepSeek-R1-Distill-Llama-70B}
        },
        xticklabel style={
            rotate=30,
            anchor=east,
            font=\small,
        },
        xmin=-0.5, xmax=1.5,
        ytick={},
        ymin=0, ymax=100,
        hide y axis,
        ymajorgrids=false,
        grid style={dashed, gray!50, line width=0.5pt},
    ]
    
    % Adding explicit horizontal grid lines
    \draw[dashed, gray!50, line width=0.5pt] (axis cs:-0.5,20) -- (axis cs:1.5,20);
    \draw[dashed, gray!50, line width=0.5pt] (axis cs:-0.5,40) -- (axis cs:1.5,40);
    \draw[dashed, gray!50, line width=0.5pt] (axis cs:-0.5,60) -- (axis cs:1.5,60);
    \draw[dashed, gray!50, line width=0.5pt] (axis cs:-0.5,80) -- (axis cs:1.5,80);
    \draw[dashed, gray!50, line width=0.5pt] (axis cs:-0.5,100) -- (axis cs:1.5,100);
    
    % Qwen3-235B
    \addplot[python_blue, fill=python_blue!30, draw=python_blue, forget plot,
              smooth, tension=0.8] table [x=x, y=y] {\pythonqwendata};
    
    \addplot[cpp_orange, fill=cpp_orange!30, draw=cpp_orange, forget plot,
              smooth, tension=0.8] table [x=x, y=y] {\cppqwendata};
    
    % DeepSeek-R1-Distill-Llama-70B
    \addplot[python_blue, fill=python_blue!30, draw=python_blue, forget plot,
              smooth, tension=0.8] table [x=x, y=y] {\pythondeepseekdistilldata};
    
    \addplot[cpp_orange, fill=cpp_orange!30, draw=cpp_orange, forget plot,
              smooth, tension=0.8] table [x=x, y=y] {\cppdeepseekdistilldata};
    
    % Draw model separators
    \addplot[black, thick] coordinates {(0, 0) (0, 100)};
    \addplot[black, thick] coordinates {(1, 0) (1, 100)};
    
    \end{axis}
    \end{scope}
  \end{tikzpicture}
  \caption{Split violin percentile distribution plots}
  \label{fig:combined_violin}
\end{figure}

Figure~\ref{fig:combined_violin} compares percentile score distributions for Python and C++ across models. Most models show bimodal patterns, with scores concentrated at the extremes (0-20 and 80-100 percentiles), indicating performance is typically either very high or very low. Gemini-2.5-flash stands out, with C++ showing mid-range density (20-40\%) and Python displaying a broader spread across all percentiles. Qwen3-235B and DeepSeek-R1-Distill-Llama-70B have balanced distributions.

\section{Limitations}
The benchmark has the following limitations:
\begin{itemize}
    \item While the benchmark supports 60 programming languages, the evaluation is only done on 2 Languages: Python and C++, due to resource constraints.
    \item Only the pass@k and percentile metrics are measured; other metrics such as execution time, memory usage, and additional performance metrics are not considered.
    \item The human baseline for a particular problem of a language is dependent on submissions made by users on \url{code.golf}. certain languages may not have  adequate submissions made by users, leading to a less accurate human baseline.
    \item Another limitations is that the number of samples generated are limited to 10 per problem, due to resource constraints.
\end{itemize}

\section{Conclusion}

This paper introduced CodeGolf Bench, a novel benchmark that significantly advances the evaluation of LLM code generation capabilities by supporting 60 programming languages far exceeding the language diversity of existing benchmarks. By leveraging the \url{code.golf} platform's dynamic ecosystem, the benchmark provides continuously updated problems and live human baselines that reflect real-time human performance, addressing a critical limitation of static benchmarks.
evaluation revealed that reasoning capabilities significantly enhance model performance in code golf tasks, particularly for languages with strict syntax requirements like C++. The substantial performance gap between reasoning and non-reasoning models (with best combined percentiles of 70.97\% versus 23.08\%) demonstrates that step-by-step logical deduction remains crucial for generating not only correct but also efficient solutions. This advantage persists across diverse problem categories, though with varying magnitudes. As LLMs continue to evolve, this benchmark will provide a dynamic, continuously updated framework for evaluating their programming capabilities against the ever-improving standard of human performance.

\textbf{Future Work} should evaluate the performance of LLMs on a wider range of all of the languages supported by the benchmark, incorporate more nuanced metrics beyond character count, and explore the relationship between solution brevity and code readability.

\clearpage

% Add these lines to configure bibliography
\bibliographystyle{plainnat}
\bibliography{references}

%%%%%%%%%%%%%%%%%%%%%%%%%%%%%%%%%%%%%%%%%%%%%%%%%%%%%%%%%%%%
\clearpage

\appendix

\section{Technical Appendices and Supplementary Material}

\begin{table}[htbp]
\caption{Benchmark comparison}
\label{tab:benchmark_comparison}
\centering
\small
\begin{tabular}{p{3cm}p{2.5cm}p{4.75cm}p{3.75cm}}
\toprule
\textbf{Benchmark comparison} & \textbf{Publicly Available Dataset} & \textbf{Publicly Available Solutions} & \textbf{Publicly Available Test Cases} \\
\midrule % Middle rule below the header
APPS & Yes & Yes & Yes \\ \hline
xCodeEval & Yes & Yes & Yes \\ \hline
CodeElo & Yes & Limited & Limited \\ \hline
LiveBench & Yes & Yes & Yes (Questions) \\ \hline
BigO(Bench) & Yes & Yes (Annotated) & Yes (Test Sets) \\ \hline
EffiBench & Yes & Yes (Canonical) & Likely \\ \hline
COFFE & Yes & Yes (Ground Truth) & Yes \\ \hline
SWE-bench & Yes & Yes (Reference) & Yes (Unit Tests) \\ \hline
HumanEval Pro & Yes & Yes (Candidate) & Likely \\ \hline
USACO Benchmark & Yes (Problems) & Yes & Yes (for recent contests) \\ \hline
\bottomrule % Bottom rule
\end{tabular}
\end{table}

\begin{table}[h]
\centering
\caption{Model Performance on C++ Code Tasks}
\label{tab:cpp_performance}
\begin{tabular}{p{5cm}p{1cm}p{1cm}p{1cm}p{1cm}p{1cm}p{1cm}p{1.5cm}}
\toprule
Model & Pass@3 & Pass@4 & Pass@5 & Pass@6 & Pass@7 & Pass@8 & Avg \%ile \\
\midrule
DeepSeek-R1 & 58.47 & 61.01 & 62.71 & 64.84 & 65.81 & 66.55 & 43.09 \\
Qwen3-235B & 52.97 & 55.94 & 57.94 & 58.90 & 60.14 & 61.22 & 42.60 \\
Gemini-2.5-pro-preview-03-25 & 47.82 & 52.98 & 55.05 & 58.47 & 58.47 & 60.19 & 24.53 \\
Gemini-2.5-flash-preview-04-17 & 54.11 & 55.30 & 52.80 & 46.43 & 40.23 & 20.77 & 38.45 \\
DeepSeek-R1-Distill-Llama-70B & 36.16 & 38.94 & 40.87 & 42.58 & 44.06 & 45.81 & 41.39 \\
Llama-4-Maverick & 46.82 & 48.84 & 51.65 & 53.53 & 55.38 & 56.35 & 40.07 \\
DeepSeek-V3-0324 & 27.83 & 28.70 & 28.70 & 31.82 & 31.82 & 31.82 & 19.13 \\
Gemma-3-27b-it & 25.09 & 25.16 & 25.19 & 25.20 & 25.21 & 25.21 & 20.66 \\
Llama-4-Scout & 12.32 & 13.41 & 14.31 & 14.80 & 15.30 & 15.48 & 16.65 \\
\bottomrule
\end{tabular}
\end{table}

\begin{table}[h]
\centering
\caption{Model Performance on Python Code Tasks}
\label{tab:python_performance}
\begin{tabular}{p{5cm}p{1cm}p{1cm}p{1cm}p{1cm}p{1cm}p{1cm}p{1.5cm}}
\toprule
Model & Pass@3 & Pass@4 & Pass@5 & Pass@6 & Pass@7 & Pass@8 & Avg \%ile \\
\midrule
DeepSeek-R1 & 58.47 & 61.01 & 62.71 & 64.84 & 65.81 & 66.55 & 67.09 \\
Gemini-2.5-pro-preview-03-25 & 61.74 & 62.61 & 63.48 & 64.35 & 64.35 & 66.09 & 68.04 \\
Llama-4-Maverick & 46.82 & 48.84 & 51.65 & 53.53 & 55.38 & 56.35 & 55.31 \\
DeepSeek-V3-0324 & 46.96 & 48.70 & 50.43 & 51.30 & 53.04 & 54.78 & 51.60 \\
Qwen3-235B & 49.13 & 52.55 & 54.97 & 57.32 & 58.61 & 59.67 & 41.64 \\
Gemini-2.5-flash-preview-04-17 & 54.11 & 55.30 & 52.80 & 46.43 & 40.23 & 20.77 & 46.42 \\
Phi-4-Reasoning-Plus & 42.95 & 46.79 & 48.81 & 50.66 & 56.94 & 56.94 & 35.21 \\
DeepSeek-R1-Distill-Llama-70B & 41.49 & 43.83 & 45.45 & 46.24 & 47.06 & 48.67 & 36.99 \\
Llama-4-Scout & 37.95 & 38.57 & 38.59 & 38.63 & 38.64 & 38.64 & 36.69 \\
Gemma-3-27b-it & 27.83 & 28.70 & 28.70 & 31.82 & 31.82 & 31.82 & 25.23 \\
\bottomrule
\end{tabular}
\end{table}

\section{Reproducibility}
\label{section:reproducibility}
Code for the benchmark is publicly available \cite{codegolfbench2025}. The gemini-2.5-pro-preview-03-25 used in the experiments is no longer available through the API. Use the gemini-2.5-pro-preview-05-06 model for future experiments. The default values for generation parameters are already set in the code.

To replicate the results, users need to sign up for an account with each API provider and set the API keys in the environment variables. Then follow the instructions in the README \cite{codegolfbench2025} to replicate the results.

\end{document}